\documentstyle[aps,manuscript,epsf]{revtex}  % DON'T CHANGE
\newcommand{\eqn}[2]{
\begin{equation}
\label{#1} #2
\end{equation}}
\newcommand{\eno}[1]{Eq.~(\ref{#1})}

\newcommand{\dd}[2]{ \frac{d #1}{d #2}}

\newcommand{\avg}[1]{\left\langle #1 \right\rangle}
\newcommand{\sbracket}[1]{\left[ #1 \right]}
\newcommand{\cbracket}[1]{\left\{ #1 \right\}}
\newcommand{\gvec}[1]{\mbox{\boldmath $#1$}}
\newcommand{\grad}[0]{\gvec{\nabla}}
\newcommand{\cH}[0]{{\cal H}}
\newcommand{\tphi}[0]{{\widetilde{\phi}}}
\newcommand{\Bar}[1]{ {\overline{ #1 } } }
\newcommand{\eps}[0]{\varepsilon}
\newcommand{\vmu}[0]{ \gvec{\mu}}
\newcommand{\tvmu}[0]{ {\widetilde{\gvec{\mu}}} }

\newcommand{\tW}[0]{ {\widetilde{W}} }
\newcommand{\hsigma}[0]{ {\widehat{\sigma}} }
\newcommand{\hg}[0]{ {\widehat{g}} }
\newcommand{\tchi}[0]{ {\widetilde{\chi}} }
\newcommand{\heps}[0]{ {\widehat{\varepsilon}} }
\newcommand{\br}[0]{{\bf r}}
\newcommand{\bh}[0]{{\bf h}}
\newcommand{\bH}[0]{{\bf H}}
\title{Vortex Glass Phase and Universal Susceptibility Variations \\
in Planar Array of Flux Lines}
\author{Terence Hwa\cite{addr} and Daniel S. Fisher}
\address{Department of Physics\\
Harvard University\\
Cambridge, Massachusetts  02138}
\date{\today}

\begin{document}
\maketitle
\widetext
\begin{abstract}
Some of the properties of the low temperature
vortex-glass phase of randomly-pinned flux lines in 1+1 dimensions
are studied. The flux arrays are found to be sensitive to small
changes in external parameters such as the magnetic field or
temperature. These effects are captured by the variations in
the magnetic response and noise, which have universal statistics
and should provide an unambiguous signature of the glass phase.
\end{abstract}
\pacs{74.60.Ge, 05.20.-y}

\narrowtext

Flux lines in a clean Type-II superconductor form an
Abrikosov lattice at low temperatures~\cite{abrikosov}.
However, the flux lattice is destroyed by random microscopic impurities
 in the material~\cite{larkin}. Recently, it has
been suggested that the disordered flux array may form
a new thermodynamic phase at low temperatures, called the ``vortex glass"
phase, in which flux lines are {\it collectively
pinned} by the impurities~\cite{mpaf,creep,ffh}.
Although glass-like behavior has been reported experimentally~\cite{expt},
a quantitative theoretical description of the vortex-glass phase
is still lacking
except in the special case of flux lines confined to a plane
(1+1 dimensions). As first shown
in Ref.~\cite{mpaf}, such a 1+1 dimensional flux array undergoes a phase
transition at a finite temperature $T_g$. Below $T_g$, the
flux array is pinned by the random impurities and forms
a glass phase. However, the properties of the glass phase
have not yet been elucidated, and a number of
contradictory results exist in the
literature~\cite{nat,toner,yedidia,balents,shapir,sudbo,ledou}.
In this paper, we analyze the vortex glass phase using
the renormalization-group method of Cardy and Ostlund~\cite{cardy}.
We find the glass phase to be characterized by
anomalous variations in the magnetic responses of the flux array,
 and extreme sensitivity to small changes in the applied field,
impurity potential, and temperature.  Such glassy behavior has been
previously conjectured for spin glasses~\cite{sg} and one flux line
\cite{zhang,mezard,dp2}, and is expected to be generic
to a wide class of randomness dominated phases.
However, the 1+1 dimensional flux array is one of the
very few systems where analytic results can be obtained.

We consider an array of flux lines confined to the $(x,z)$-plane,
with an applied field ${\bf H} = H_z {\bf \hat{z}}$ and
repulsive  interactions which we model by linear
elasticity~\cite{nelson,hnv}.
Impurites yield a random potential $V(x,z)$.
Labeling the transverse displacement of the $n$-th line by $r_n(z) =
(n - \phi_n(z)/2\pi)/\rho$, where $\rho \sim H_z$
is the average line density,
we can describe the large scale fluctuations
of the flux array by the  Hamiltonian
\cite{mpaf,nat,toner,hnv}
\eqn{H} {  \cH = \int dx dz \cbracket{\frac{\kappa}{2} \sbracket{
(\partial_x\phi)^2 + (\partial_z\phi)^2} - V(x,z)\partial_x\phi
- 4\pi\rho V \cos\sbracket{2\pi \rho x +\phi } }, }
where $\phi(x,z)$ is the coarse-grained displacement field.
In \eno{H}, the $x-$ and $z-$ dimensions have been rescaled
to make the quadratic part isotropic. The elastic coefficient
$\kappa \sim (d\rho/dH)^{-1}$ is weakly temperature dependent.
The cosine term in \eno{H} comes from the invariance of the system to an
overall shift in the labeling index $n$ of the lines. It
 picks out the {\it discrete} nature of the
flux lines and is crucial to the formation of a glass phase.

Upon renormalization, one generates a term of the
 form $V'(x,z)\partial_z\phi$, which randomly
biases the local {\it tilt} of the flux lines.
It is found that the variances of $V$ and $V'$ are renormalized
in the same way, so that the inherent spatial anisotropy and
frustration present in \eno{H}
disappear at large length scales. It is then more
convenient to work with the isotropic Hamiltonian,
\eqn{Hphi}
{ \cH[\phi] =  \int_\br \cbracket{ \frac{\kappa}{2}
(\grad\phi)^2 - \vmu \cdot\grad \phi
- W(\phi(\br),\br) },}
where $\br =\{x,z\}$, $\vmu(\br) = V(\br) {\bf \hat{x}}
 + V'(\br) {\bf \hat{z}}$ is gaussian distributed
with mean zero and variance of the component $\mu_i$
\begin{equation}
\Bar{\mu_i(\br)\mu_j(\br')} = \sigma \delta_{ij}
\delta^2(\br-\br'),\label{mu}
\end{equation}
and $W(\phi,\br) \propto \cos[\phi(\br)-\beta(\br)]$ is a random potential,
 describing the effect of a random phase  $\beta(\br)$, with
\begin{equation}
\Bar{W(\phi,\br) W(\phi',\br')}
= 2g \cos[\phi-\phi']  \delta^2(\br-\br').\label{W}
\end{equation}
Denoting the bare parameters by the subscript $0$, we have $g_0 \sim
\sigma_0 \rho^2$. A renormalization-group analysis~\cite{cardy,sr}
yields the recursion relations under a change
of scale by $b= e^l$,
\begin{eqnarray}
  &\dd{\kappa}{l} &= 0, \label{rg1} \\
  &\dd{\sigma}{l} &= A g^2, \label{rg2} \\
  &\dd{g}{l} &= \eps g - C g^2. \label{rg3}
\end{eqnarray}
The coefficients $A$ and $C$ are cutoff dependent, however
the ratio $A/(\kappa C)^2 = 2\pi +O(\eps)$ is {\it universal}.
Eqs.~(\ref{rg2}) and (\ref{rg3}) are valid
to leading order in $g$, which will be sufficient provided
\begin{equation}
\eps \equiv 2 - \frac{T}{2\pi \kappa} \label{eps}
\end{equation}
is small.
This is true even though $\sigma$ can flow to large values by
Eq.~(\ref{rg2}), because
the random potential $\vmu$ in \eno{Hphi} can be shifted away by the
transformation $\phi'(\br) = \phi(\br) - u(\br)$, with
$ \kappa \nabla^2 u = \grad \cdot \vmu$,
regardless of the magnitude of $\vmu$.
The resulting potential  $W'(\phi',\br) = W(\phi'+u,\br)$
 has the same statistics
as $W(\phi,\br)$ since the latter is uncorrelated in $\br$.
Consequently, the flow of $g$ cannot be affected by $\vmu$.
A similar use of the statistical symmetry of $\cH$ shows that
the result that $\kappa$ is unrenormalized in Eq.~(\ref{rg1})
is {\it exact}~\cite{sr,schulz}.

Clearly, $\eps=0$ is a special point; it defines a critical temperature
$T_g = 4\pi\kappa$ through Eq.~(\ref{eps}). For $T>T_g$
where $\eps < 0$, $g$
renormalizes to zero and at long scales the system is described by the
gaussian part of the Hamiltonian (\ref{Hphi}) with a {\it finite}
renormalized $\sigma$.
This is the flux liquid phase with the disorder causing
only short wavelength modifications  of the pure system~\cite{nelson}.
But for $T<T_g$ where $\eps > 0$,
there is  a nontrivial phase controlled by a fixed line $g^*(T)$,
with $\sigma$ renormalizing as in Eq.~(\ref{rg2}).
Close to the transition, we have $g^* \approx \eps/C \propto (T_g - T)$,
and on scale $L$,
$\sigma(L) \approx A (g^*)^2 \log \rho L$. This is
a vortex glass phase~\cite{note1}.

The existence of a perturbatively accessible
 fixed line allows us to study the vortex glass phase
quantitatively. The nonrenormalization of $\kappa$ implies a simple form for
 the mean square thermal fluctuations
 of $\phi$, i.e., $\Bar{\avg{[\phi(\br)-\phi(\br')]^2}_c}
= T/(2\pi \kappa) \log[\rho |\br-\br'|]$ the same as
in the absence of randomness~\cite{schulz}.
The glass phase is instead distinguished by more strongly divergent
static distortions. For example, the mean square (thermally averaged)
 displacement is $\Bar{\avg{\phi(\br)-\phi(\br')}^2}\approx
\eps^2 \log^2[\rho |\br-\br'|]$ due to
the logarithmic divergence of $\sigma$~\cite{sr}.
However, this  is not a unique feature of a  glass
phase, as systems with long-range correlated $\vmu$'s
can also give rise to anomalous mean square displacement
even if $g=0$,
in which case the system is harmonic and trivial.

We therefore consider other quantities
whose  behavior is unique to a glass phase.
We first study the magnetic response of the flux array.
 We change the applied external field by
an amount $\delta {\bf H} = \delta H_x {\bf \hat x}
+ \delta H_z {\bf \hat z}$,
which tends to compress and/or rotate the flux array.
For an isotropic system, the Hamiltonian becomes
\eqn{Hh}{\cH_\bh[\phi] = \cH[\phi] - \int_\br
\> \bh \cdot \grad \phi,} where
$\bh = (\delta H_z {\bf \hat x} + \delta H_x {\bf \hat z})\Phi_0/(8\pi^2)$,
 $\Phi_0$ being the magnetic flux quantum.
The change in the flux density is $\langle\partial_x
\phi \rangle/(2\pi)$, and in the ``tilt angle'' is  $\langle\partial_z
\phi \rangle/(2\pi\rho)$.
The linear response on which we focus is
$\chi_{i,j} \equiv \frac{\partial}{\partial h_i}
\langle \partial_j \phi \rangle$. For the
isotropic system~(\ref{Hphi}), we have $\chi_{ij} = \chi \delta_{ij}$,
and the magnetic permeability is just $[\Phi_0^2/(16\pi^3)] \ \chi$.
Simple rescaling yields a similar result for the anisotropic system.

Consider the high temperature phase where discreteness is irrelevant,
i.e., $g=0$.  Then the last term in (\ref{Hh})
can be simply shifted away by the transformation
\begin{equation}
\phi'(\br) = \phi(\br) - \bh\cdot\br/\kappa, \label{tilt}
\end{equation}
 yielding a  free energy
\begin{equation}
 F(\bh) = -\frac{h^2}{2\kappa} L^2 - \frac{1}{\kappa}
\int_\br \> \bh \cdot \vmu,\label{F0}
\end{equation}
and hence a response $\Bar{\chi}  = 1/\kappa$. Since the random part of
$F(\bh)$ is linear in $\bh$, the response will be {\it sample
independent} as in a pure system, with $\Bar{(\Delta\chi)^n} = 0$
for $n>2$
where $\Delta\chi \equiv \chi - \Bar{\chi}$.
This  is solely a consequence of the quadratic nature of the
Hamiltonian with $g= 0$.

The magnetic response in the low temperature phase, where
the random phase term in~(\ref{Hphi}) is relevant, is much more interesting.
Since  the transformation (\ref{tilt}) does not change the statistics
of the Hamiltonian $\cH$~\cite{sr,schulz}, except
for generating an extra
quadratic term as in Eq.~(\ref{F0}),
the quenched-averaged free energy $\Bar{F(\bh)}$
is the {\it same} as for $g=0$.
 Thus, $\Bar{\chi} = 1/\kappa$ independent of $g$.
Furthermore, the average of higher order
nonlinear susceptibilities $\Bar{(\Delta\chi)^n}$
all vanish due to the statistical symmetry~\cite{schulz}.
Thus average response functions are identical in the glass
and liquid phases.  This result has led
some to doubt mistakely even the existence
of the glass phase~\cite{sudbo}. However we will show that the
 glassy effects are manifested
in sample-to-sample variations of the susceptibility and
its extreme sensitivity to small perturbations.

Let us compute the effect of the random potential
$W$  on the susceptibility variation, perturbatively at first.
After the transformation (\ref{tilt}),
the correlations between the free energy at two different fields
$\bh_1$ and $\bh_2$ can be calculated to the lowest order in $g$.
For $\Delta F(\bh) \equiv F(\bh) - \Bar{F(\bh)}$, we have
\begin{equation}
\Bar{\Delta F(\bh_1)\Delta F(\bh_2)}
= 2 g (\rho L)^{-\frac{T}{2\pi \kappa}}
 \int_\br \cos\sbracket{(\bh_1-\bh_2)\cdot \br/\kappa} \label{dF}
\end{equation}
for a system of size $L\times L$, with the $(\rho L)^{-T/(2\pi \kappa)}$
factor arising from averaging over thermal fluctuations. Differentiating
with respect to $\bh_1$ and $\bh_2$
leads to nontrivial sample-to-sample variations of
the magnetic susceptibility, with variance
\begin{equation}
\Bar{(\Delta\chi)^2}
=  \frac{D g}{\rho^2\kappa^4} \cdot (\rho L)^\eps \label{dx1}
\end{equation}
 to first order in $g$,
with $D$ being a sample-geometry dependent coefficient.

For $T>T_g$ (i.e., $\eps <0$), Eq.~(\ref{dx1}) gives the form of
approach to the asymptotic liquid phase where
$\Bar{(\Delta\chi)^2}=0$ as discussed above.
For $T < T_g$, $\Bar{(\Delta\chi)^2}$ diverges since
$\eps > 0$, indicating the failure of the small-$g$ expansion.
Eq.~(\ref{dx1}) does, however, suggest the form of the correct
behavior: the term $g (\rho L)^\eps$ should just be replaced by the
renormalized $g_R(L)$. Explicit computation shows that this is
indeed the case. For large systems in the vortex glass phase,
$g_R(L) \to g^*$ hence we obtain a fractional variance
\begin{equation}
\frac{ \Bar{(\Delta\chi)^2} } { \Bar{\chi}^2 } \approx
\frac{ D g^*}{ \rho^2 \kappa^2} = \widehat{D} \eps \label{dx}
\end{equation}
with $\widehat{D}$ a {\it universal} geometry and boundary condition
dependent coefficient, that is
independent of nonmeasurable bare parameters such as $g_0$.
For an isotropic square sample with periodic boundary condition,
we have $\widehat{D} \approx 8\pi/5$.
The large sample-to-sample variations of $\chi$ indicate
that the vortex glass phase is radically different from the fluid
phase \cite{note2}.
The size-independent variations of $\chi$ in the vortex glass phase
 are reminiscent of
``universal conductance fluctuations" in disordered metals~\cite{ucf}.

Experimentally, variations of $\chi$  may be obtained
by measuring the magnetic response of {\it one} sample at different
applied fields $\bH$. It will be particularly convenient to
keep $|\bH|$ and $T$ fixed, and follow the
response as the {\it direction}  ${\bf \hat{H}}$ is changed.
The variance $\Bar{(\Delta\chi)^2}$ only  depends on $T$ and
  $|\bH|$ (through $\kappa$).
Then as ${\bf \hat{H}}$ is changed, say by rotating a sample
in a fixed field, it effectively samples  different
``realizations" of the random potential, drawn from the {\it same}
distribution since systems with different field directions
${\bf \hat{H}}$ are {\it statistically equivalent}~\cite{note3}.
For a system of size $L\times L$,
the free energies and hence the susceptibilities become uncorrelated
if  ${\bf \hat{H}}$ is changed by an angle much greater than
 $ (\rho L)^{-1}$ as can be guessed from  \eno{dF}
with $\kappa \sim \Phi_0 H/\rho$.
In the glass phase, we thus expect
to obtain a wildly varying susceptibility $\chi({\bf \hat{H}})$,
whose precise form  is a property of the specific sample,
but with universal statistics, in particular,
$\Bar{(\Delta\chi)^2} / \Bar{\chi}^2$.
Alternatively, one could monitor  the magnetic noise as a function
of ${\bf \hat{H}}$. This should
 exhibit universal variations like $\chi({\bf \hat{H}})$,
since the two are related by the fluctuation-dissipation
relation.
The susceptibility variations at a fixed $T$ and $|\bH|$
provide ``magnetic finger prints" of the glassy flux phase.
The reproducibility of the magnetic finger print for the same sample
under identical conditions provides a probe  of thermal equilibrium
on long scales: Only samples small enough to equilibrate fully (see below)
 will show reproducible behavior.

{}From the above discussion, it is evident that the equilibrium
 state of the flux array
depends sensitively on small changes in the applied field.
As argued in Refs.~\cite{sg} and~\cite{dp2} on general grounds,
a wide class of random systems can
exhibit such sensitivity to small changes of a variety of
parameters such as a field or temperature.
Large variations resulting from small changes in the
random potential $V(\br)$ have been
studied numerically by Zhang~\cite{zhang} for
 a single flux line.
In the remainder of this paper, we analyze explicitly the effect of
such a small change in the random potential for the 1+1 dimensional
 flux array.
Sensitivity of the array to small temperature changes can be analyzed
similarly. We merely quote the analogous result for this somewhat more
complicated case.

We consider two noninteracting flux arrays, $\phi(\br)$ and $\tphi(\br)$,
in two {\it different} realizations of the random potential,
 $\{\vmu(\br),W(\phi,\br)\}$
and $\{\tvmu(\br),\tW(\phi,\br)\}$ respectively.
We take the random potentials to be statistically equivalent but slightly
different from each other, so that $\Bar{\tvmu(\br)\tvmu(\br')}$
is given by \eno{mu} and $\Bar{\tW(\phi,\br)\tW(\phi',\br')}$
 given by \eno{W}.
However the cross-correlators are
\begin{eqnarray}
& & \Bar{\mu_i(\br)\widetilde{\mu}_j(\br')} = \hsigma
\delta_{ij}\delta^2(\br-\br'), \label{tmu} \\
& & \Bar{W(\phi,\br)\tW(\phi',\br')}
= 2 \hg \cos[\phi-\phi'] \delta^2(\br-\br'), \label{tW}
\end{eqnarray}
with the bare values $\hsigma_0 < \sigma_0$ and $\hg_0 < g_0$.
The renormalization group recursion relations Eqs.~(\ref{rg1}) --
(\ref{rg3}) must be unchanged  as the systems are uncoupled.
However, the cross correlations renormalize as
\begin{eqnarray}
  &\dd{\hsigma}{l} &= A \hg^2, \label{rg4} \\
  &\dd{\hg}{l} &= \heps \hg - C g \hg, \label{rg5}
\end{eqnarray}
with
$ \heps \equiv \eps + (\hsigma-\sigma)/(2\pi \kappa^2)$.

To investigate the effect of weak de-correlation of the random potentials
in the glass phase, we linearize the recursion relations around the
vortex glass fixed point $g^*$. For small de-correlation $\delta_\sigma =
\sigma - \hsigma \ll 1$ and $\delta_g = g-\hg \ll 1$, we have
\begin{eqnarray}
 &\dd{\delta_\sigma}{l} &= 2A g^* \delta_g,  \\
 &\dd{\delta_g}{l} &= \frac{1}{2\pi\kappa^2} g^* \delta_\sigma
\end{eqnarray}
This flow has one positive eigenvalue
\begin{equation}
\lambda_\delta \approx g^* \sqrt{A /\pi\kappa^2}
= \sqrt{2}\eps. \label{eigen}
\end{equation}
Therefore,
 infinitesimally small
 de-correlations in the bare random potential grow under renormalization.
On long scales $L \gg L_\delta \sim \delta^{-1/\lambda_\delta}$, with
$\delta$ a linear combination of $\delta_\sigma$ and $\delta_g$, $\hg(L)$
vanishes and $\hsigma(L)$ saturates.
The two systems then appear substantially
different and will have essentially
{\it independent} susceptibilities, with
$\Bar{\Delta\chi(L)\Delta\tchi(L)} \to 0$ for large $L$.  There will,
however, be residual cross correlations associated with the finite
renormalization of $\hsigma$.

These effects can best be probed by changing the  temperature of one sample
 slightly by $\delta T$. The same exponent $\lambda_\delta$
in Eq.~(\ref{eigen})
controls the crossover, and for system sizes $L\gg L_\delta \sim
(\delta T)^{-1/\lambda_\delta}$, $\chi(T)$ and $\chi(T+\delta T)$
will be roughly independent. If $\delta T \ll T_g - T$, the temperature
dependence of $\chi$ will probe statistically similar variations
of $\chi$ as did the field direction dependence of $\chi({\bf \hat{H}})$.

Physically the source of the sensitivity to $\bH$ and $T$ changes are
quite different. The former is due to  the changes in  mean position
Of the lines while the latter is more subtle:
It is caused by the {\it entropic}
contributions to the free energy, which drastically changes
 the effective random
potential on long scales. Although this has been predicted for a variety of
random systems~\cite{sg,dp2}
 and supported by numerical and approximate renormalization
group calculations, this to our knowledge, is the first time an analytic
calculation has yielded the hypersensitivity to temperature changes.

We close with a comment  on dynamics:   Recently, a number of authors
have claimed that free energy barriers in this system grow as various
powers of $\log L$~\cite{nat,toner,ledou}. An explicit dynamic
renormalization-group calculation~\cite{shapir,goldschmidt}
found that the dynamic exponent $z \approx 2 + 1.8\eps$ for $T\le T_g$,
yielding a nonlinear resistivity, $d{\cal E}/dJ \sim J^{0.9 \eps}$,
where ${\cal E}$
is the EMF generated by a uniform current $J$ applied perpendicular to the
$(x,z)$-plane. However, because the 1+1 dimensional vortex glass phase
is controlled by a finite temperature fixed line rather than a zero
temperature fixed point, the barriers are not well defined by the form
${\cal E}(J)$ found. In the limit $T\to 0$, however, one finds~\cite{dsf}
$z \sim 1/T$, which can then be correctly interpreted
as barriers growing as $\log L$.

The dynamics can also be used to probe the length dependence of $\chi$.
At finite frequency, $\omega$, scales of size $l_\omega \sim \omega^{-1/z}$
 are probed. Since the susceptibility $\chi(\omega)$ for each correlation
volume $l_\omega^2$
will be essentially independent, the variations in $\chi(\omega)$ for
a sample of size $L\times L$ will be
$\Delta \chi(\omega) \sim l_\omega/L \sim \omega^{-1/z}/L$,
crossing over to the static result only when $l_\omega \sim L$.

In this paper, we have analyzed some of the glassy properties
 of randomly pinned flux arrays confined to a plane.
In the vortex glass phase, the magnetic susceptibility is found to be
strongly dependent on the external field, temperature and
specific sample, exhibiting
variations with universal statistics.

We wish to thank L. Balents, M. Kardar,
T. Nattermann, and D. R. Nelson for useful discussions. TH is grateful
to the hospitality of NORDITA where part of this work was completed.
This research is supported by NSF through Grants No.~DMR-91-06237
and DMR-91-15491, and by the Harvard University
Materials Research Laboratory.


\begin{references}
\bibitem{note1} In Ref.~\cite{nat}, it was shown that $\eps > 0$ for
all $T$ in a free-fermion model of flux lines which forbids line crossing.
An alternative derivation~\cite{hnv} using a more realistic (soft) repulsive
potential allows both signs of $\eps$.
\bibitem{note2} Note that $\chi$ only probes the system
on the {\it largest} scale, $L$. Therefore, it
does not get a $\log L$ divergence as,
for example, in the mean square displacement of $\phi$, which contains
contributions from all scales up to $L$.
\bibitem{note3}The exact statistical equivalence is obtained only
with periodic boundary conditions.
In general the statistics will depend on the sample shape. However for tilt
angles which are small but still much larger than $(\rho L)^{-1}$, the
different
field directions will be statistically equivalent.
\end{references}
\end{document}